# Electromagnetic transients and failed upward leaders observed during lightning activity in an onshore wind farm

Franjo Vukovic, Bozidar-Filipovic-Grcic, Nina Stipetic, Bojan Franc

*Abstract*—At a wind farm in Croatia, lightning activity is monitored across the entire site using a lightning location system, and on a single wind turbine equipped with a Rogowski-coil-based current measurement system and a high-speed camera, all independently GPS-synchronized. In addition to recording lightning flash currents on the monitored turbine, the system is frequently triggered by electromagnetic disturbances caused by nearby lightning flashes. These include direct flashes to two neighboring turbines that share the same cable connection to the substation and have interconnected grounding systems with buried bare conductor, as well as cloud-to-ground flashes to soil near cable routes, where the resulting electromagnetic fields couple onto the cables, causing surges to propagate to the monitored turbine. The camera occasionally captures failed upward connecting leaders from the monitored turbine during these lightning events. This paper presents three cases of flashes to two neighboring wind turbines and two cases of cloud-to-ground flashes to nearby soil, all of which induced electromagnetic transients that propagated to the monitored turbine. Failed upward connecting leaders were observed in some of these cases. This paper provides observational analysis, providing Rogowski measurements of electromagnetic disturbances and failed leader currents, complemented by high-speed camera and lightning location system data.

## I. Introduction

Wind energy sector continues to experience significant growth, as highlighted in the *Global Wind Report 2024* [1], which reported a record installation of 117 GW in 2023, predominantly from onshore wind turbines (WTs). Onshore WTs are relatively tall, isolated from other structures and potentially located at high altitudes in mountainous or hilly regions to maximize wind energy extraction. However, these factors significantly increase their exposure to lightning flashes. Given the rapid expansion of wind farm (WF) installations, the frequency of lightning-induced issues is expected to rise. This challenge is further exacerbated when the grounding conditions at the site are poor. Consequently, the design of lightning protection and grounding systems must be optimized to mitigate the adverse direct and indirect effects of lightning strikes and ensure the reliable operation of WTs in electromagnetic environment during lightning activity.

Lightning activity generates electromagnetic transients that propagate through the electrical system, either from direct strikes to a WT or indirect strikes to the nearby soil, where electromagnetic coupling induces transients in collector cables. In both cases, these surges travel through the wind farm's electrical network. Current research focuses on time-domain simulations of electromagnetic transients in WFs [2]-[14] and the coupling of lightning electromagnetic pulses with power cables [15]-[18].

To improve transient response, systems of specific WTs are interconnected with buried bare or insulated conductors [19] along the routes of power collection cables. Additionally, multiple turbines sometimes share the same underground collector cable connection to the substation, further influencing transient propagation. Since WTs are interconnected through cables and buried conductors, lightning-induced surges can propagate across the cable network to other WTs as well as the power grid [14], [20].

In a WF lightning observatory located in southern Croatia, lightning activity is monitored using a lightning location system (LLS), two Rogowski coils, and a high-speed (HS) camera [21]-[24]. The Rogowski coils are installed on a single WT, which is also monitored by an HS camera positioned at a nearby substation. Similar measurements of lightning currents using Rogowski coils were conducted on WTs in Japan. These investigations include the NEDO (New Energy and Industrial Technology Development Organization) project [25], MHI (Mitsubishi Heavy Industries, Ltd.) studies [26], as well as investigations by J-POWER (Electric Power Development Co., Ltd.) and CRIEPI (Central Research Institute of Electric Power Industry) [27]. Additionally, there is extensive experience in using Rogowski coils on tall and static towers, such as the Santis Tower [28], CN Tower [29], Tokyo Skytree [30], the Windmill and its protection tower in Japan [31], Sentech Tower [32], and Canton Tower [33].

Onsite measurement experience at a Croatian WF shows that when one of the two neighboring WTs is directly struck or when a nearby cloud-to-ground (CG) flash to soil occurs, one or both Rogowski coils register electromagnetic disturbances.

This work was supported by the Croatian Science Foundation (HRZZ, DOK-2021-02), European Regional Development Fund within project DESMe, KK 01.1.1.07.0028 and nowcast GmbH (providing LLS data).

F. Vukovic, B. Filipovic-Grcic, N. Stipetic and B. Franc are with University of Zagreb Faculty of Electrical Engineering and Computing, Unska 3, 10000 Croatia (e-mail: franjo.vukovic@fer.hr, bozidar.filipovic-grcic@fer.hr, nina.stipetic@fer.hr, bojan.franc@fer.hr).

Paper submitted to the International Conference on Power Systems Transients (IPST2025) in Guadalajara, Mexico, June 8-12, 2025.

These disturbances, caused either by direct hits to neighboring turbines or by CG flashes to soil whose electromagnetic impulses couple with power cables, produce surges that propagate to the monitored WT. In such events, one or both Rogowski coils detect the disturbances, which are recorded by the acquisition system if they exceed the trigger threshold. The HS camera, triggered by the current measurement system, captures footage of these events, providing either a clear view of CG flashes occurring within the frame or partial changes in luminance across the frame during flashes to neighboring turbines, as the camera is only focused on the monitored turbine. Notably, during some of the events described, whether involving neighboring WTs or CG flashes to soil, the camera also captured failed upward connecting leaders (UCLs) from the monitored turbine.

This paper presents an observational study of these events, analyzing data from Rogowski coil measurements, HS camera recordings, and LLS detections of flashes and associated electromagnetic disturbances. The propagation of electromagnetic disturbances in the WF lightning observatory was briefly mentioned in [22].

## II. Monitoring lightning activity with Rogowski coils, HS camera and LLS: experience with measurements up to date

Figure 1 shows a map view of the WF, illustrating the northwest section of the 110/20 kV substation and comprising seven WTs. These WTs, each rated at 3 MW, have a tower height of 132 m and 49 m-long blades. While only seven WTs are shown on the map, an additional batch of seven WTs is located west of the substation, outside the displayed area. The map highlights a smaller circle for each WT, representing their equivalent collection areas with a radius three times the WT height (396 m), as per IEC 61400-24 [19]. A lightning strike identified by the LLS within the equivalent collection area is considered to have struck the respective WT. LLS data is used for continuous monitoring of lightning activity, generating weekly statistics and spatially correlating lightning flashes with specific WTs. This correlation is performed for each WT. The orange triangles indicate the actual locations of the WTs, while the green triangles represent corrected positions, adjusted to compensate for the LLS location bias (toward the southwest), ensuring accurate spatial correlation of lightning strokes detected by the LLS.

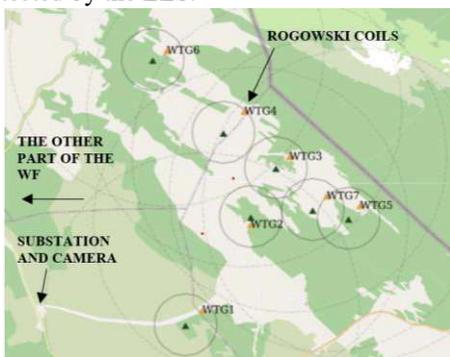

Fig. 1. Map view of the WF showing the section northwest of the 110/20 kV substation, which includes 7 WTs.

In addition to monitoring the activity with the LLS, lightning observations are conducted using a current and HS camera measurement system on Wind Turbine Generator (WTG) No. 4 (WTG4), which is located at the highest altitude of around 1250 m. Current measurements have been ongoing since July 2022, while HS camera measurements started in April 2023. The current measurements are based on two Rogowski coils, each connected via a coaxial cable to an integrator:

- a high-frequency (HF) sensor with a 1 MHz bandwidth and a range of ±250 kA,
- and a low-frequency (LF) sensor with a 10 kHz bandwidth and a range of ±12.5 kA.

The HF and LF sensors include a Rogowski coil, a coaxial cable, and an integrator. Both Rogowski coils are wrapped around the base of the WT tower, approximately 1 meter above the ground, and are connected to coaxial cables that are routed inside the WT tower to an electrical cabinet where integrators are. The acquisition system for current measurement is also located inside the electrical cabinet of WTG4. The HS camera is located approximately 3.5 km away at the substation, where it is connected to a central server that manages its triggering and data archiving process. The camera is set to be triggered by the current measurement system as the current measurement system on WTG4 and central server at a substation are interconnected via optical cables.

Up until now, more than 100 lightning current waveforms have been collected over the two and a half years of monitoring. For the HS camera, which has been operational for a year and a half, around 20 videos have been captured where the lightning channel of flashes that struck WTG4 are clearly visible. During the winter season, the visibility of the channel is poorer. Additionally, at least as many videos have been collected where the lightning channel cannot be seen due to poor visibility conditions. However, luminosity changes across the entire frame can still be registered, allowing for the extraction of luminosity profiles of those flashes and correlation with other systems.

Furthermore, there are numerous HS camera recordings of nearby lightning activity, where the lightning flash is either clearly visible in the frames or appears only as a general luminosity change across the entire frame. These cases include nearby lightning activity between clouds behind WTG4, nearby CG flashes to soil, and strikes to one of the two neighboring WTs, WTG3 or WTG6 (Figure 1), which share the same 20 kV cable connection to the local substation. The earthing systems of these WTs are also interconnected with a buried bare conductor. WTG6 is located 1250 m away from WTG4, while WTG3 is approximately 750 m away, both measured as air distances.

During the cases described above, some camera recordings capture UCLs initiated from WTG4 that ultimately fail to connect. For instance, during a lightning flash to nearby soil, as the downward leader progresses toward the ground, UCL can be observed originating from the WTG4 blade in several frames. This leader persists until the downward leader successfully connects with the ground, at which point the UCLs disappear.

Electromagnetic disturbances associated with these events are detected by the LF sensor, HF sensor, or both, depending on whether they are intense enough to trigger the acquisition system. If a UCL is present, then the LF sensor captures the associated continuous current. In the following section, a few examples of nearby lightning events are analyzed using Rogowski measurements, HS camera recordings, and LLS data of the flashes and their corresponding electromagnetic disturbances.

It is important to note that the temporal precision of the current measurement system is up to at least a microsecond, while the LLS data used in the analysis has a temporal precision of up to a millisecond. Therefore, since all events occur within or around the 1 ms range, more precise GPS timestamps would be more suitable for a detailed analysis.

### III. RESULTS

#### A. Lightning Flash to WTG3 with Failed UCL initiated from WTG4

On July 6th, 2023, the HS camera captured a lightning event involving WTG4. The camera acquisition settings were configured to 5000 fps (period: 200 μs), with an exposure time of 6 μs and a resolution of 640×600 pixels. The total recording time was 1.146 s, including 0.5982 s of pre-trigger time.

Figure 2 shows two frames from the initial stages of the event, spaced approximately 1 ms apart. These frames clearly illustrate the progression of the downward leader from the cloud and the upward connecting leader initiated from the WT blade, moving towards each other. Figure 3 (left) represents the moment when a sudden increase in luminance occurred on the right side of the HS camera frame as the downward leader approached WTG4. After this moment, the leader began to withdraw and gradually faded away, ultimately failing to connect to WTG4. According to the HS camera's timestamp, the sudden and intense increase in luminance occurred at 16:11:56.975 UTC. This coincided with the moment the LLS registered a stroke hitting the WTG3 to the right of WTG4 (from the camera's perspective) and another stroke hitting WTG6 (left from the camera's perspective). Figure 3 (right) shows the exact locations of the two strokes as determined by the LLS, marked with red crosses. According to LLS, the WTG6 stroke occurred at 16:11:56.975 UTC, while the WTG3 stroke occurred 16:11:56.974 UTC.

The camera was triggered by the current measurement system. In this instance, the system potentially registered a continuous current associated with an initiated UCL, as shown in Figure 4. Immediately afterward, an electromagnetic transient was observed, caused by a lightning stroke to WTG3. The timing of this stroke, as determined by the LLS, is marked by the vertical line in the figure. Only the timing of the stroke that impacted WTG3 is plotted. No records from the HF sensor are available for this event.

Typically, electromagnetic transients caused by strokes to WTG3 and WTG6 are also captured by the HF sensor, which then provides a detailed view of their waveforms, provided the transients exceed the trigger threshold. This will be demonstrated through the following two examples.

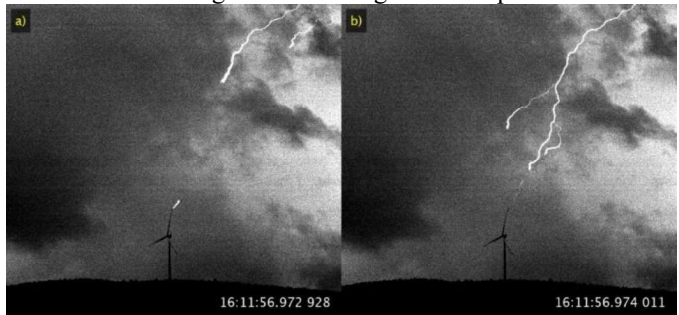

Fig. 2. Stepped downward leader and UCL progression for July 6th, 2023, lightning flash event.

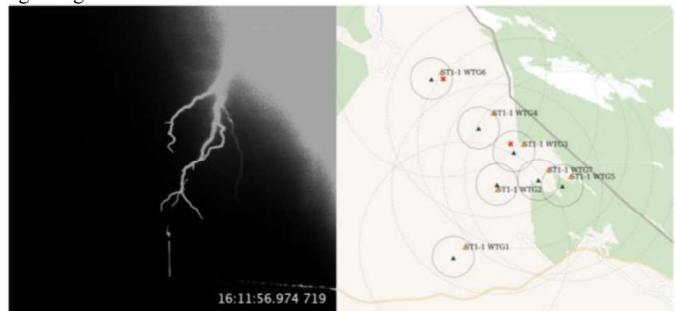

Fig. 3. HS camera frame when the luminance suddenly increased (left); Lightning flash activity recorded by LLS in that area at that exact time (right).

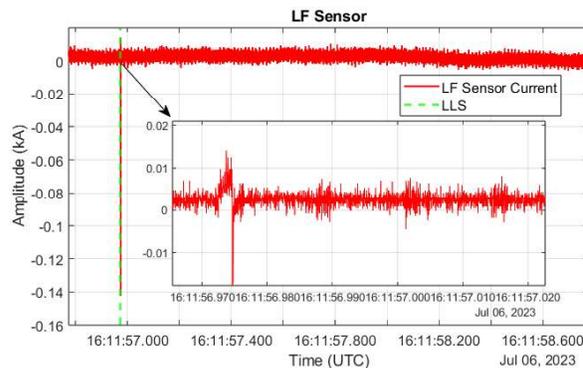

Fig. 4. LF sensor record of July 6th, 2023, lightning flash event: potential continuous current immediately followed by electromagnetic disturbance.

#### B. Lightning flashes to neighboring WTG6 and WTG3

This section examines two examples of lightning flashes, one to WTG3 and one to WTG6. The first example involves a lightning flash to WTG6 on December 11th, 2022, while the second highlights a lightning flash to WTG3 on January 21st, 2023. Lightning flashes striking WTG3 and WTG6 generate electromagnetic transients that propagate along the grounded cable sheaths and buried bare conductors interconnecting the grounding systems. These transients ultimately reach WTG4, where both HF and LF sensors detect the resulting electromagnetic disturbances.

Figure 5 shows the LLS spatial data associated with December 11th, 2022, lightning flash to WTG6, which includes 15 successive strokes located in the collection area of the WT (marked with blue cross), and data associated with the January 21st, 2023, lightning flash to WTG3, which includes 4 successive strokes located in the collection area of the WT (marked with yellow cross).

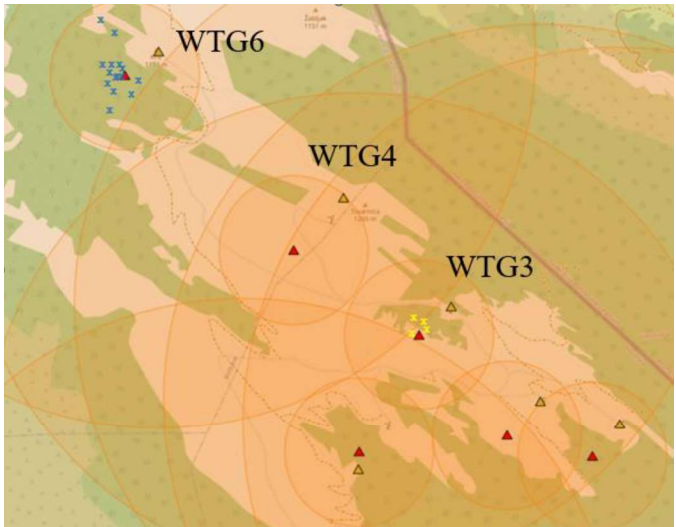

Fig. 5. December 11[th], 2022, lightning flash to WTG6: 15 strokes (marked with blue cross); Jan 21[st], 2023, lightning flash to WTG3: 4 strokes (marked with yellow cross).

The lightning flash to WTG6 on December 11[th], 2022, occurred at approximately 16:21:46 UTC. Figure 6 presents data collected by the LF and HF sensor at WTG4, highlighting the electromagnetic disturbances generated by this event. The figure includes a 2-second LF sensor measurement capturing the entire event (Figure 6a), several 1.5-millisecond HF sensor measurements, each corresponding to the transient caused by an individual stroke (Figure 6b), and a combined plot of LF and HF sensor measurements annotated with vertical lines marking the timings of strokes detected by the LLS (Figure 6c). The figure is zoomed in on the y-axis to enhance the visibility of the LF sensor disturbances. Figure 6 further illustrates the HF sensor records by plotting them together, aligned to time zero, to emphasize the similarity of the electromagnetic disturbances in shape and duration. After a stroke occurs on WTG6, an associated electromagnetic disturbance is detected by the current measurement system (HF and LF sensors) in about a millisecond.

The LF sensor captured all the electromagnetic disturbances that reached WTG4, as clearly shown in Figure 6c, where almost every local peak on the LF sensor corresponds to the time the LLS detected a stroke to WTG6. The first two peaks on the LF sensor are probably not associated with disturbances caused by strokes and they may instead be linked to the continuous current from UCL. However, there is no camera data for this case, as the camera was not yet been installed.

Out of the total 15 CG strokes, the HF sensor picked up electromagnetic transients associated with six. The remaining transients likely did not have sufficient intensity upon reaching WTG4 to trigger the measurement system. As shown in Figure 6, there are seven HF sensor records in total; however, the first record contains a disturbance not associated with any stroke detected by the LLS. This is further confirmed by Figure 7, which illustrates that this first disturbance captured by the HF sensor differs significantly from the other electromagnetic disturbances in terms of duration and shape.

The remaining transients generally follow a consistent pattern, both in shape and duration (several hundred microseconds).

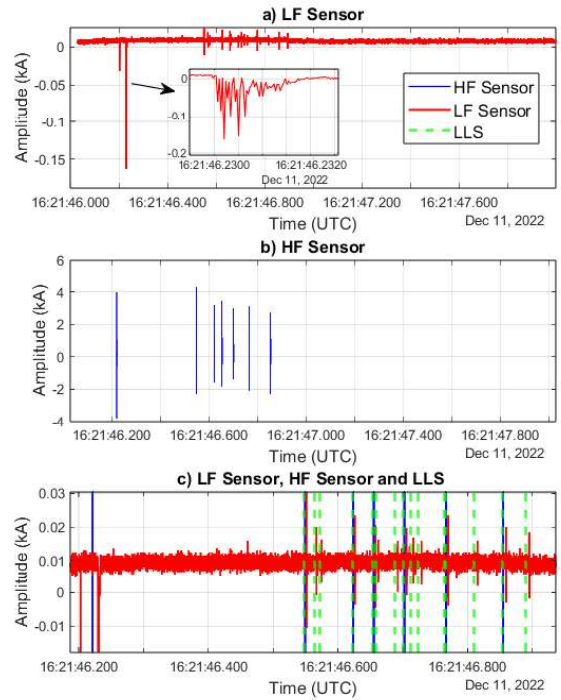

Fig. 6. Electromagnetic transients at WTG4 caused by December 11[th], 2022, lightning flash to WTG6, as recorded by the LF and HF Rogowski coils.

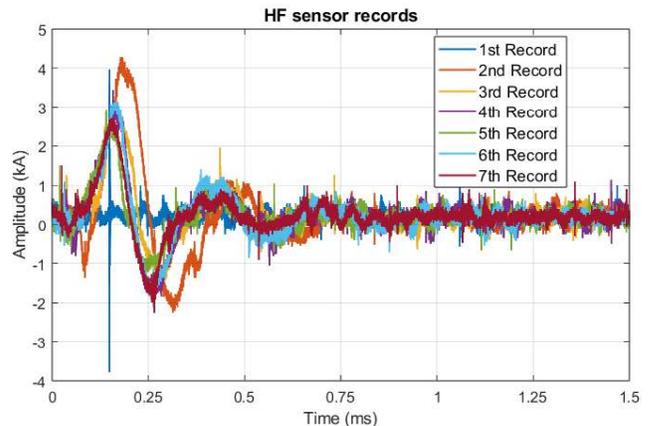

Fig. 7. HF sensor records of electromagnetic transients caused by the December 11[th] lightning flash to WTG6, aligned to time zero.

The lightning flash to WTG3 on January 21[st], 2023, occurred at approximately 05:33:59 UTC. Figure 8 presents data collected during this event, including a combined plot of HF sensor measurements annotated with vertical lines marking the timings of strokes detected by the LLS (Figure 8a) and a plot of all HF sensor records aligned to time zero to emphasize their similarities in waveform shape and duration (Figure 8b).

For this event, no LF sensor record is available as it was not triggered. The HF sensor, however, captured all electromagnetic disturbances associated with each of the four strokes detected by the LLS. There is approximately a 1 ms offset between the stroke and the corresponding disturbance detected at WTG4, except for the last case, where they are almost perfectly aligned. This discrepancy arises from the LLS data being precise only to the millisecond, while the propagation of transients occurs much faster. More precise LLS timestamps would therefore be needed for a more accurate analysis.

As shown in Figure 8, the disturbances during the initial phase exhibit similar oscillatory behavior, which gradually attenuates overtime. Notably, the third (orange) and fourth (purple) HF sensor records contain additional oscillations occurring later in the event.

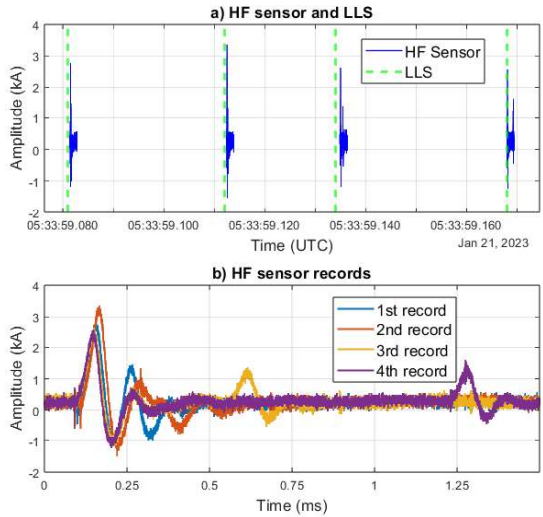

Fig. 8. Electromagnetic transients at WTG4 caused by the January 21st, 2023, lightning flash to WTG3, as recorded by the LF and HF Rogowski coils.

*C. Nearby CG Flashes to soil and Associated Electromagnetic disturbances*

This section presents two examples of lightning flashes to nearby soil and the associated electromagnetic disturbances induced on cables and/or grounding systems, which propagated to WTG4. The examples include data from the current measurement system, HS camera recordings, and LLS data for the lightning flashes on September 3rd, 2024, around 13:55:35 UTC and April 25th, 2024, around 15:59:21 UTC. In both cases, the LF sensor measurements indicate the presence of continuous currents. However, the presence of continuous current associated with UCL is further confirmed only for the September 3rd event, as several HS camera frames clearly show the UCL initiated from the WT blade. Figure 9a illustrates the LLS spatial data associated with the CG lightning flash on September 3rd, 2024, while Figure 9b shows a similar CG flash that occurred on April 25th, 2024. In both cases, the strokes are numbered from 1 to 5.

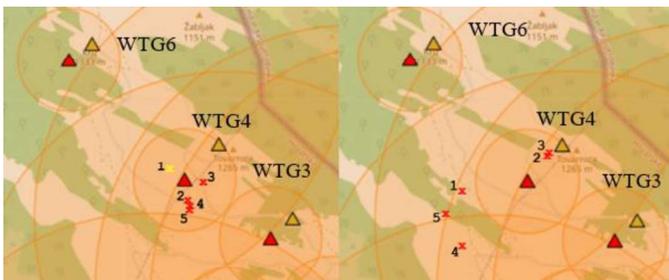

Fig. 9. September 3rd, 2024, lightning flash (Figure 9a, left): April 25th, 2023, lightning flash (Figure 9b, right).

First, the September 3rd, 2024, lightning event will be analyzed. According to Figure 9a there were five discharges detected with the first one being classified as cloud-to-cloud (yellow cross). Figure 10 shows specific camera frames corresponding to moments around the times when the LLS detected strokes. In the top right corner of Figure 10 there are labels that associate each frame with specific LLS numbering of the stroke used in Figure 10a. Camera captured in total six strokes, one more than LLS as LLS missed the last one. Furthermore, the first stroke, which was categorized as cloud-to-cloud, appears to have hit the ground, as indicated by the first camera frame. From the frames the strokes used the same channel and discharged at the same ground strike point.

Figure 11 provides information about the electromagnetic disturbances caused by the lightning flash. Specifically, Figure 11a shows the LF sensor measurement of the event, indicating that a UCL occurred on the WT blade. However, the UCL failed, as indicated by the abrupt cessation of continuous current. Figure 12 displays two non-successive camera frames capturing the moment when the UCL was visible. Notably, the UCL and the associated continuous current faded as the flash made its initial connection with the soil. It should be noted that the UCL was visible in more frames than the two shown in Figure 12, but its appearance in the additional frames was much less distinct.

Figure 11b illustrates the LF and HF sensor measurements overlaid with vertical lines representing the LLS-detected stroke timings. As shown in Figure 11b, each vertical line corresponding to an LLS-detected stroke aligns with a local disturbance captured by the LF sensor. The HF sensor detected electromagnetic disturbances associated with two out of the five strokes detected by the LLS: the first and third stroke, following the numbering used in Figure 9a.

These transients are presented in greater detail in Figure 11c, where the corresponding, originally 9-ms long, HF sensor records are displayed relative to each other, enabling better comparison of the waveform shapes of the measured disturbances. The transients exhibit a similar overall waveform shape. Compared to flashes occurring directly on neighboring wind turbines, the transients observed here have a shorter duration (around 0.1 ms) and are not as oscillating.

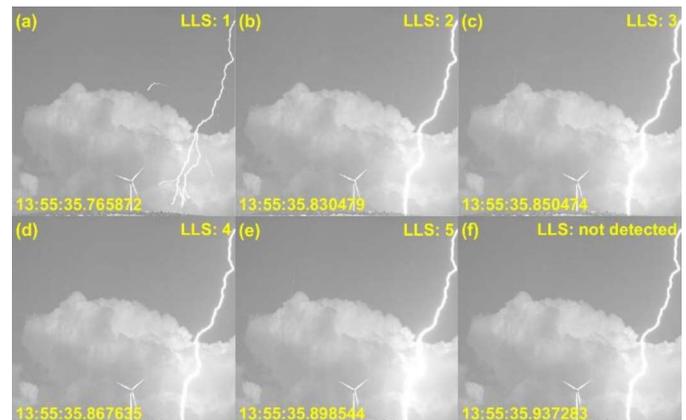

Fig. 10. September 3rd, 2024, lightning flash: HS camera captured failed UCL and six strokes in total.

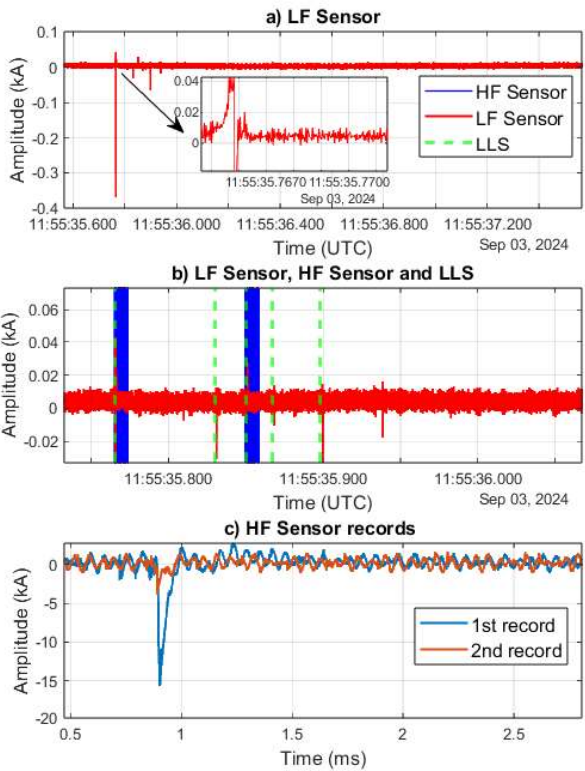

Fig. 11. Continuous current associated with failed UCL and electromagnetic transients at WTG4 caused by nearby CG flash to soil on September 3$^{rd}$, 2024.

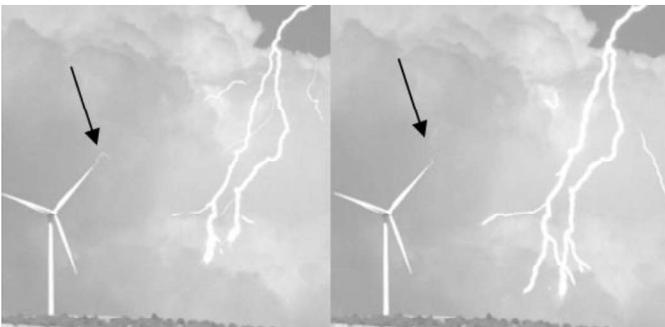

Fig. 12. Non-successive HS camera frames before the first connection, in which the UCL initiated from the blade, are visible.

Now, the lightning event occurred on April 25$^{th}$, 2023, will be analyzed. According to Figure 9b, five discharges were detected by the LLS. Despite the clear spatial separation of the discharges, they likely belong to a single flash, as the different ground strike points of the same flash can be several kilometers apart. Figure 13 shows specific HS camera frames corresponding to moments around the times when the LLS detected strokes. In the top right corner of each frame in Figure 13, labels are added to associate each frame with the numbering of strokes detected by the LLS in Figure 9b.

The first camera frame (a) corresponds to the first stroke detected by the LLS. For this stroke, neither the camera nor the measurement system registered any associated disturbance, suggesting that the stroke likely occurred farther away. Additionally, the camera detected one stroke not recorded by the LLS, occurring between strokes 4 and 5, as shown in frame (e). As indicated by the camera frames, these discharges belong to two different flashes: one occurring behind the WT (captured in frames (b) and (c)) and the other in front of the WT (captured in frames (d), (e), and (f)).

Figure 14 provides information collected by the current measurement system installed on WTG4 for this event. Specifically, Figure 14a shows the LF sensor measurement of the entire event, while Figure 14b overlays the HF sensor measurements onto the LF sensor data, with vertical lines indicating the timing of LLS-detected strokes. Figure 14c illustrates the HF sensor records aligned relative to each other for waveform comparison.

As indicated by the LF sensor measurement, it is likely that a UCL was launched during this event. However, it cannot be observed in the camera frames due to cloud cover obstructing a clear view of the WT blades. The timing of the measured continuous current corresponds approximately to the time of the frame (b) in Figure 13. In this frame, the cloud covering the upper part of the WT is clearly visible.

There are two HF sensor records – one is associated with the first flash behind the WT (camera frame (b) in Figure 13) and the other is associated with the first stroke in front of the WT (camera frame (d) in Figure 13).

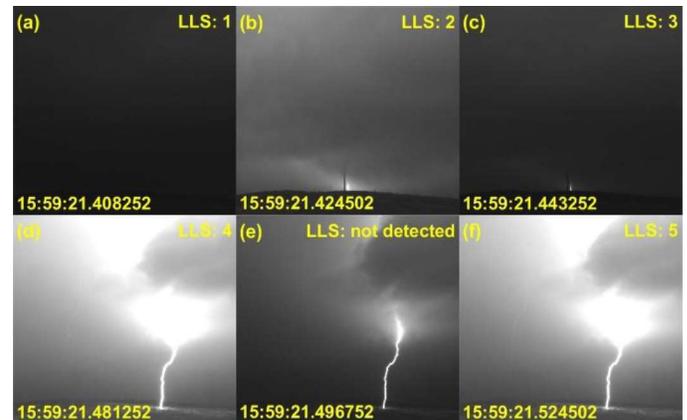

Fig. 13. April 25$^{th}$, 2023, lightning flash: HS camera captured five strokes in total.

In Figure 14c, these events are presented in greater detail, where the corresponding 9-ms long HF sensor records are displayed relative to each other. The electromagnetic transient associated with HF sensor record 1, which corresponds to the stroke behind the WT, does not exhibit oscillations. In contrast, HF sensor record 2, representing the stroke in front of the WT shows more oscillatory behavior. Similar to the previous CG flash event to soil, the durations of these disturbances are shorter than those observed when one of the neighboring WTs is directly struck.

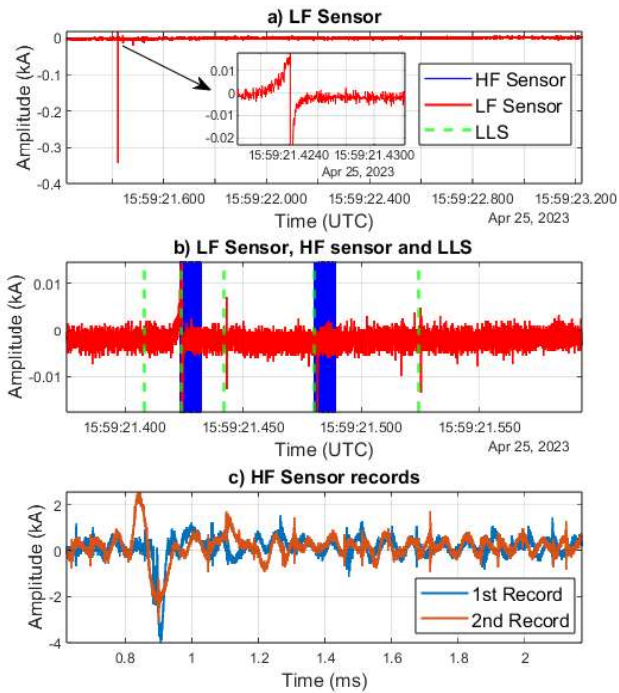

Fig. 14. Continuous current associated with failed UCL and electromagnetic transients at WTG4 caused by nearby CG flash to soil on April 25, 2023.

## IV. Conclusions

In a WF in southern Croatia, lightning is monitored using an LLS, a Rogowski-coil-based current measurement system, and an HS camera. The current measurement system is installed on WTG4, the highest-altitude turbine in the WF (approximately 1250 m above sea level). The HS camera, located at a nearby 110/20 kV substation, is focused on WTG4 and is triggered by the current measurement system. WTG4 shares collector cables and interconnected grounding systems with WTG6 (1250 m northwest) and WTG3 (750 m southeast). Direct lightning flashes to WTG6 and WTG3 cause electromagnetic disturbances that propagate to WTG4, triggering the current measurement system and the HS camera. Furthermore, the electromagnetic field generated by CG flashes to soil couples with the collector cables and grounding system, resulting in surges that propagate across the WF and are detected by the current measurement system installed on WTG4.

During direct flashes to nearby WTs or the soil, WTG4 occasionally initiates a UCL. These UCLs, often visible in HS camera recordings, appear as brief continuous currents in LF sensor measurement data. However, they frequently fail to connect, as the lightning discharge strikes another WT or the local soil.

This paper discusses direct lightning events involving WTG3 and WTG6 as well as several nearby CG flashes to soil. In one event, a lightning flash struck WTG3 while UCL was initiated from WTG4. HS camera footage showed the UCL and a lightning branch moving toward each other, ultimately failing to connect as the flash struck WTG3. The LF sensor recorded both the continuous current associated with the UCL and the electromagnetic transient from the stroke, while LLS confirmed the flash's location within WTG3's collection area. The HF sensor did not detect the stroke, likely because its intensity was insufficient to trigger the system.

Two earlier events occurred before the HS camera was installed. In the first, a lightning flash struck WTG6, with LLS recording 15 strokes. Both HF and LF sensors detected electromagnetic disturbances associated with these strokes. In the second event, a flash struck WTG3, with LLS recording four strokes, all of which were detected by the HF sensor. The transients in both cases exhibited oscillatory waveforms with durations of several hundred microseconds.

Two nearby CG flash events to the soil are analyzed, both involving HF and LF sensor measurements, HS camera footage, and LLS data. In one event, a UCL initiated from WTG4 and was captured in the HS camera footage, with the corresponding continuous current detected by the LF sensor. In the other event, continuous current was measured; however, cloud cover obscured the WT blades in the camera footage, preventing confirmation of a UCL. According to the HF sensor records, the transients associated with these disturbances are less oscillatory and shorter in duration compared to those caused by direct strikes to neighboring WTs.

The reason transients recorded are shorter in duration and less oscillatory in cases of CG flashes to soil, compared to direct flashes to neighboring WTs, lies in the difference in coupling mechanisms. In general, when a lightning flash directly strikes one of the neighboring WTs, the transient propagates directly along the cable sheath, as the sheath is connected to the grounding system. This results in less attenuation and more oscillatory transients. On the other hand, when a CG flash strikes the soil and couples indirectly with the cable sheath, the transient experiences significantly more attenuation, leading to fewer oscillations and shorter-duration transients. The above-described comparison is valid for the same source that produces the transients, meaning the lightning flash has the same amplitude and steepness. In our case, however, it is difficult to draw straightforward conclusions because the source amplitude and steepness differ for each analyzed scenario.

Future work will be focused on installing additional field sensors and cameras to collect more information about lightning activity throughout the entire WF area. LLS data, field sensor measurements and measured lightning current waveforms could be used to perform accurate EMT simulations for investigating electromagnetic transients caused by direct or nearby lightning strikes to WF. Obtaining more details about measured lightning current parameters and lightning activity could be used for WF maintenance and improving lightning/overvoltage protection of WTs.

## V. References

[1]  Global Wind Energy Council (GWEC), Global Wind Report 2024, 2024.

[2] R. Alipio, M. Guimarães, L. Passos, D. Conceição, M.T.C. de Barros, Ground Potential Rise in Wind Farms due to Direct Lightning, *Electric Power Systems Research* 194 (2021) 107110. https://doi.org/10.1016/j.epsr.2021.107110.

[3] R. Alipio, M.T. Correia de Barros, M.A.O. Schroeder, K. Yamamoto, Analysis of the lightning impulse and low-frequency performance of wind farm grounding systems, *Electric Power Systems Research* 180 (2020) 106068. https://doi.org/10.1016/j.epsr.2019.106068.

[4] K. Yamamoto, T. Matsumoto, T. Fukunaga, R. Alipio, Measured and simulated impulse responses of the grounding systems of a pair of wind-turbines connected by a buried insulated wire, *Electric Power Systems Research* 213 (2022) 108726. https://doi.org/10.1016/j.epsr.2022.108726.

[5] W.C. da Silva, W.L.M. de Azevedo, A.R.J. de Araújo, J.P. Filho, Full-wave electromagnetic analysis of lightning strikes to wind farm connected to medium-voltage distribution lines, *Electric Power Systems Research* 223 (2023) 109597. https://doi.org/10.1016/j.epsr.2023.109597.

[6] W.C. Da Silva, W.L.M. De Azevedo, J.L.A. D'Annibale, A.R.J. De Araújo, J.P. Filho, Transient Analysis on Wind Farms with Interconnected Grounding Systems Located on Frequency-Dependent Soils, in: *2023 IEEE Power & Energy Society General Meeting (PESGM)*, IEEE, 2023: pp. 1–5. https://doi.org/10.1109/PESGM52003.2023.10252629.

[7] R. Shariatinasab, B. Kermani, J. Gholinezhad, Transient modeling of the wind farms in order to analysis the lightning related overvoltages, *Renew Energy* 132 (2019) 1151–1166. https://doi.org/10.1016/j.renene.2018.08.084.

[8] Md.R. Ahmed, M. Ishii, Electromagnetic analysis of lightning surge response of interconnected wind turbine grounding system, in: *2011 International Symposium on Lightning Protection*, IEEE, 2011: pp. 226–231. https://doi.org/10.1109/SIPDA.2011.6088443.

[9] M.E.M. Rizk, A. Ghanem, S. Abulanwar, A. Shahin, Y. Baba, F. Mahmood, I. Ismael, Induced Electromagnetic Fields on Underground Cable Due to Lightning-Struck Wind Tower, *IEEE Trans Electromagn Compat* 65 (2023) 1684–1694. https://doi.org/10.1109/TEMC.2023.3303280.

[10] S. Sekioka, H. Otoguro, T. Funabashi, A Study on Overvoltages in Windfarm Caused by Direct Lightning Stroke, IEEE Transactions on Power Delivery 34 (2019) 671–679. https://doi.org/10.1109/TPWRD.2018.2883910.

[11] S.M.A. Hosseini, A. Mohammadirad, A.A. Shayegani Akmal, Surge analysis on wind farm considering lightning strike to multi-blade, *Renew Energy* 186 (2022) 312–326. https://doi.org/10.1016/j.renene.2021.12.061.

[12] N. Malcolm, R. Aggarwal, Analysis of transient overvoltage phenomena due to direct lightning strikes on wind turbine blade, in: *2014 IEEE PES General Meeting | Conference & Exposition*, IEEE, 2014: pp. 1–5. https://doi.org/10.1109/PESGM.2014.6938780.

[13] C. Yeung, J. Wang, M. Zhou, J. Cao, Y. Ding, L. Cai, Y. Fan, Q. Zhou, J. Wang, W. Zhao, Transient induced response on cables by lightning electromagnetic pulse with different experimental conditions, *Electric Power Systems Research* 238 (2025) 111031. https://doi.org/10.1016/j.epsr.2024.111031.

[14] N. Malcolm, R.K. Aggarwal, Transient overvoltage study of an Island wind farm, in: *2012 47th International Universities Power Engineering Conference (UPEC)*, IEEE, 2012: pp. 1–6. https://doi.org/10.1109/UPEC.2012.6398450.

[15] L. Boussayoud, B. Nekhoul, Transient electromagnetic lightning disturbances induced in the collector cables of a wind farm, *Electric Power Systems Research* 241 (2025) 111345. https://doi.org/10.1016/j.epsr.2024.111345.

[16] M.E.M. Rizk, F. Mahmood, M. Lehtonen, E.A. Badran, M.H. Abdel-Rahman, Investigation of Lightning Electromagnetic Fields on Underground Cables in Wind Farms, *IEEE Trans Electromagn Compat* 58 (2016) 143–152. https://doi.org/10.1109/TEMC.2015.2493206.

[17] S. Sekioka, H. Otoguro, T. Funabashi, A Study on Overvoltages in Windfarm Caused by Direct Lightning Stroke, *IEEE Transactions on Power Delivery 34* (2019) 671–679. https://doi.org/10.1109/TPWRD.2018.2883910.

[18] X. Li, J. Wang, Y. Wang, Y. Fan, B. Zhang, S. Wang, M. Zhou, L. Cai, Lightning transient characteristics of cable power collection system in wind power plants, IET Renewable Power Generation 9 (2015) 1025–1032. https://doi.org/10.1049/iet-rpg.2014.0449.

[19] IEC, Wind energy generation systems-Part 24: Lightning protection (No. 61400-24), 2019.

[20] W. Maallem, B. Nekhoul, S. Kaouche, Realistic modeling of direct lightning strike on a wind farm: grounding systems considerations, *Electrical Engineering* 106 (2024) 3269–3282. https://doi.org/10.1007/s00202-023-02149-y.

[21] F. Vukovic, B. Filipovic-Grcic, N. Stipetic, B. Franc, Installation and initial measurement results of the Rogowski-coil-based wind turbine lightning current waveform measurement system, *Electric Power Systems Research* 238 (2025) 111061. https://doi.org/10.1016/j.epsr.2024.111061.

[22] F. Vukovic, B. Filipovic-Grcic, N. Stipetic, B. Franc, D. Brändlein, Correlation of Prototype Lightning Current Waveform Measurement System With LLS Data, in: *37th International Conference on Lightning Protection (ICLP)*, Dresden, 2024.

[23] F. Vukovic, V. Milardic, B. Filipovic-Grcic, N. Stipetic, B. Franc, D. Milos, Incorporating a High-speed Camera in the Lightning Current Measurement System for Wind Turbines, in: 2023 4th *International Conference on Smart Grid Metrology (SMAGRIMET)*, IEEE, 2023: pp. 1–6. https://doi.org/10.1109/SMAGRIMET58412.2023.10128662.

[24] F. Vukovic, V. Milardic, D. Milos, B. Filipovic-Grcic, N. Stipetic, B. Franc, Development and laboratory testing of a lightning current measurement system for wind turbines, *Electric Power Systems Research* 223 (2023) 109572. https://doi.org/10.1016/j.epsr.2023.109572.

[25] M. Saito, M. Ishii, Characteristics of return strokes associated with upward lightning flashes observed in winter, in: *Proc. 34th International Conference on Lightning Protection (ICLP)*, IEEE, 2018, pp. 1–5, https://doi.org/10.1109/ ICLP.2018.8503444.

[26] Y. Ueda, M. Fukuda, T. Matsushita, S. Arinaga, N. Iwai, K. Inoue, Measurement experience of lightning currents to wind turbines, Ratio 4 (2007) 2.

[27] A. Asakawa, T. Shindo, S. Yokoyama, H. Hyodo, Direct Lightning Hits on Wind Turbines in Winter Season Lightning Observation Results for Wind Turbines at Nikaho Wind Park in Winter, *IEEJ Transact. Electr. Electron. Eng. 5* (2010) 14–20, https://doi.org/10.1002/tee.20487.

[28] F. Rachidi, M. Rubinstein, Säntis lightning research facility: a summary of the first ten years and future outlook, *E & i Elektrotechnik Und Informationstechnik 139* (2022) 379–394, https://doi.org/10.1007/s00502-022-01031-2.

[29] S. Kazazi, P. Liatos, A. Hussein, Simultaneous recording of CN tower lightning current and channel luminosity, in: *2014 IEEE 27th Canadian Conference on Electrical and Computer Engineering (CCECE)*, IEEE, 2014, pp. 1–6, https://doi. org/10.1109/CCECE.2014.6901035.

[30] T. Miki, A. Kudo, M. Saito, Overview of lightning current observation at Tokyo Skytree, in: *2023 XXXVth general assembly and scientific symposium of the international union of radio science (URSI GASS)*, IEEE (2023) 1–4, https://doi.org/10.23919/URSIGASS57860.2023.10265489.

[31] X. Wang, D. Wang, J. He, N. Takagi, Characteristics of electric currents in upward lightning flashes from a windmill and its lightning protection tower in Japan, 2005–2016, *J. Geophys. Res. Atmospheres 126* (2021), https://doi.org/10.1029/ 2020JD034346.

[32] J.R. Smit, H.G.P. Hunt, C. Schumann, The Johannesburg Lightning Research Laboratory, Part 1: characteristics of lightning currents to the Sentech Tower, *Electr. Pow. Syst. Res. 216* (2023) 109059, https://doi.org/10.1016/j. epsr.2022.109059.

[33] L. Chen, W. Lyu, Y. Ma, Q. Qi, B. Wu, Y. Zhang, Q. Yin, H. Liu, Y. Zhang, S. Chen, X. Yan, S. Du, Return-stroke current measurement at the Canton Tower and preliminary analysis results, *Electr. Pow. Syst. Res. 206* (2022) 107798, https:// doi.org/10.1016/j.epsr.2022.107798.